\documentclass{article}
\usepackage{spconf,amsmath,graphicx}
\usepackage{mathtools}
\usepackage{amssymb}
\usepackage{amssymb}
\usepackage{textcomp}
\usepackage{graphicx}
\usepackage{graphics}
\usepackage{epsfig}
\usepackage{epstopdf}
\usepackage{float}
\usepackage{color}
\usepackage[figuresright]{rotating}
\usepackage{listings}
\usepackage{etoolbox}
\usepackage{amsmath, amsthm, amssymb, bm}
\usepackage{mathtools, cuted}
\usepackage{amsmath}
\usepackage{qtree}
\usepackage{tikz}
\usepackage{subfigure}
\usepackage{adjustbox}

\title{Hand Gesture Recognition Using Ultrasonic Waves}
\name{Mohammed H. AlSharif \qquad Mohamed Saad \qquad \textbf{Advisor:} Tareq Y. Al-Naffouri}

\address{ EE Department, King Abdullah University of Science \& Technology, Thuwal, Saudi Arabia\\
\{mohammed.alsharif, mohamed.saadeldin, tareq.alnaffouri\}@kaust.edu.sa}

\begin{document}
\ninept
\maketitle
\begin{abstract}
 This paper presents a new method for detecting and classifying a predefined set of hand gestures using a single transmitter and a single receiver utilizing a linearly frequency modulated ultrasonic signal. Gestures are identified based on estimated range and received signal strength (RSS) of reflected signal from the hand. Support Vector Machine (SVM) was used for gesture detection and classification. The system was tested using experimental setup and achieved an average accuracy of $88\%$.
\end{abstract}

\section{Introduction}
\label{sec:intro}

With nowadays spread usage of electronic devices and the advances in human-computer interface technology, gesture control is becoming an active research area. This technology has applications in consumer electronics, medical care, advertisement and many other applications\cite{Gest2015Market}.
There are two main approaches for touchless gesture recognition systems; active and passive. Active approach requires the user to wear or hold a device such as data gloves or infrared sensor. Passive appraoch doesn't require user physical contact \cite{rautaray2015vision}. A popular passive gesture recognition approach is based on cameras. An alternative attractive approach is based on utilizing reflected ultrasonic waves from the user's hand.
Compared to camera-based approach, ultrasonic insures user privacy, insensitive to illumination changes, and has lower power consumption and computational complexity.\\
In \cite{gupta2012soundwave} Gupta et al. developed an algorithm to track Doppler shift caused by the moving hand using laptops in different environments. Kalgaonkar et al. in \cite{kalgaonkar2009one} developed a simple device based on Doppler effect to recognize one-handed gesture using low-cost ultrasonic transmitter and receivers.\\
In this paper we presents a novel hand gesture recognition system based on ultrasonic ranging. The system is of low complexity and can detect five types of gestures using a single ultrasonic transmitter and a single ultrasonic receiver.\\
In the second section of this paper, system parameters and signal design are presented. In the third section, signal processing techniques used are explained. Fourth section presents gesture detection and classification using Support Vector Machine (SVM). The fifth section describes experimental setup and shows the experimental results. Final section concludes the paper.

\section{System Parameters and Signal Design}
\label{sec:format}
\begin{figure}[h!]
\centering
\includegraphics[width = 85mm, height= 35mm]{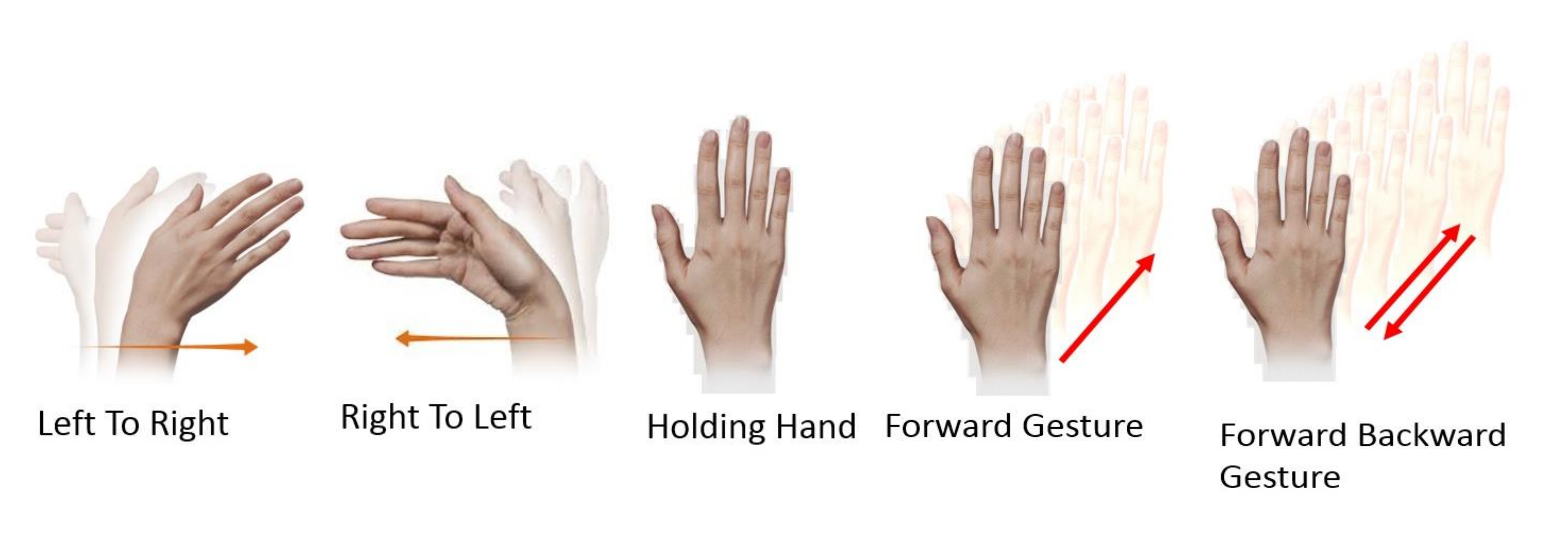}
\caption{The predefined set of hand gestures}
\label{gest}
\end{figure}

\begin{figure}[!h]
\centering
\includegraphics[width = 85mm, height= 30mm]{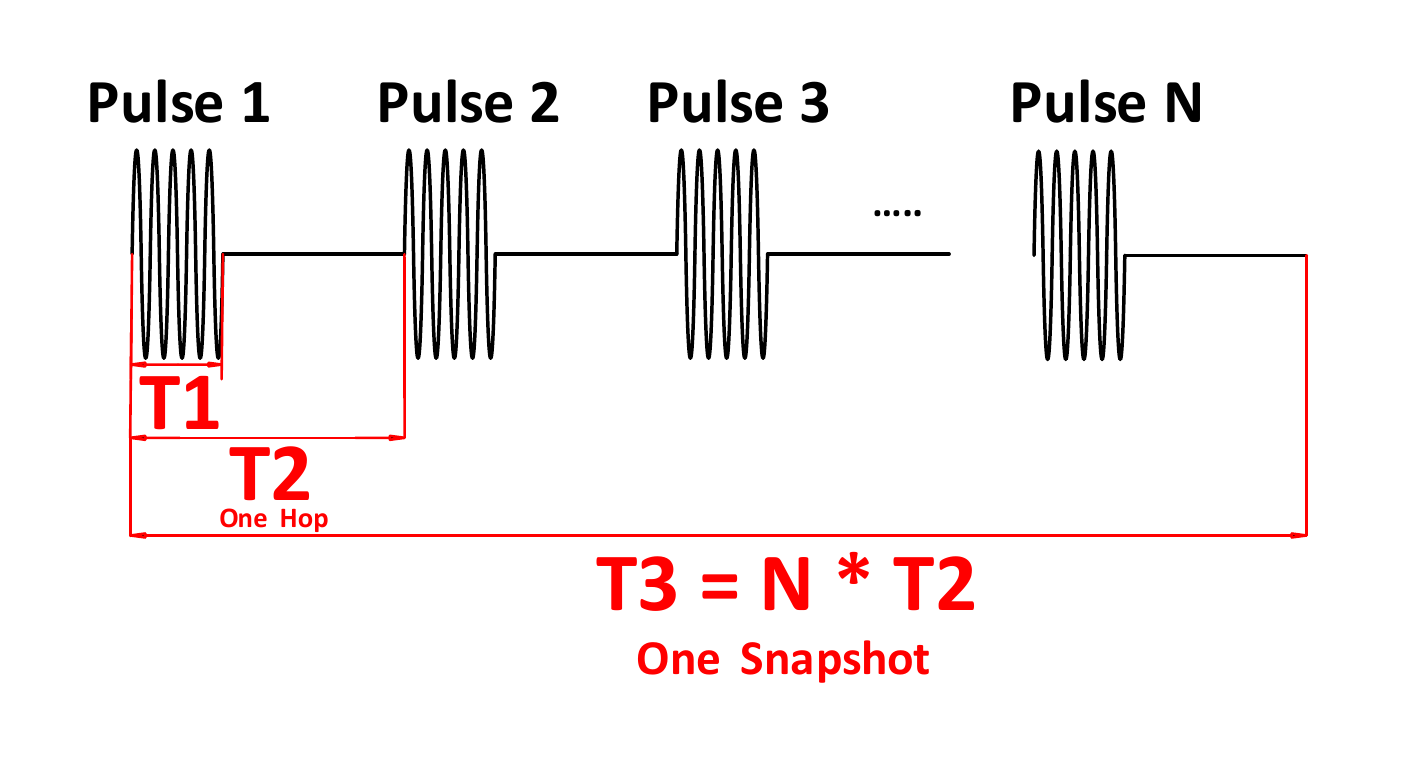}
\caption{Transmitted train of pulses for one block}
\label{train}
\end{figure}

The system aims to detect five types of gestures, shown in Figure \ref{gest}, performed by a  single hand moving in a range of 10-50 cm in depth and 40 cm horizontally with a speed not exceeding 1 m/s. 
The transmitted signal is a train of pulses designed to satisfy the requirements of the system. Figure \ref{train} shows the transmitted signal and highlights the parameters $T_1$, $T_2$ and $T_3$. The transmitter and receiver are separated by a distance of 1.1 cm. In addition to the reflected signals, a self-interference signal is received directly from the transmitter. It's a signal design objective to reduce the overlap between the reflected  and the self-interference signals. The nearest reflected signal from the hand arrives in 0.6 ms. Therefor, the value of $T_1$ was sat to 0.5 ms. The farthest reflected signal from the hand arrives in 2.9 ms. However, to help with reducing the effect of multi-path reverberations, the value of $T_2$ was sat to 5 ms.
To increase SNR, a block of N pulses is processed for each estimation. The value of N should be small enough to assume static hand over the time period $T_3 = N \times T_2$. Therefore, the value of N was sat to 4.\\
 Odd indexed pulses are up-Chirps and even indexed pulses are down-Chirps. This provides orthogonality between consecutive pulses. Each pulse has a central frequency at 38.8 KHz and bandwidth of (38.8 $\pm$ 3.5) KHz.

\section{Signal Processing Techniques}
\label{sec:majhead}
Processing the reflected signals aim is to estimate the range and RSS. The reflected signal from hand can be modeled as:
\begin{equation}
y[n] = \alpha x[n-d] + w[n]
\end{equation}
Where $\alpha$ is attenuation factor, $x[n]$ is the transmitted signal, $d$ is TOF from the transmitter to the hand then back to receiver, and $w[n]$ is AWGN.  
Cross-correlation is applied between the transmitted and received signal to estimate TOF ($d$) and RSS ($\alpha$). De-cluttering is used to remove the self-interference and other unwanted reflections from static and slowly moving objects. Peaks associated with static or slowly moving objects will always appear at the same delay (TOF) over all cross-correlation frames. Subtracting the previous cross-correlation frames from the current one will remove them. If we denote the cross correlation vector $v_i$ where $i$ refers to the index of the processed block, and let the cluttering factor be $c $, then the output of the de-clutter is given by:
\begin{equation}
\bm{v}_{\hat{i}} = \bm{v}_i - \left( c^i \bm{v}_1 + c^{i-1} (1-c) \bm{v}_2 + ...  + c^0 (1-c) \bm{v}_{i-1} \right)
\label{declutter}
\end{equation}
Where $c \in [0,1]$.
Cross correlating and de-cluttering produce Motion Frames in which peaks positions relates to the current range of gesturing hand and their amplitude indicates RSS. 
\section{Gesture Detection and Classification }
\label{sec:print}
\begin{figure}[!h]
\includegraphics[width=1\linewidth]{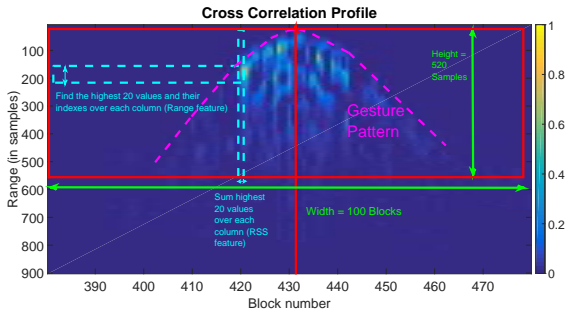}
\caption{RSS and Range vectors generation}
\label{rss_range}
\end{figure}

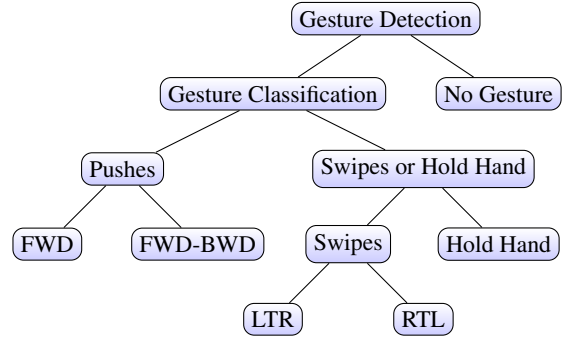
\begin{figure}[h!]
\begin{tikzpicture}[level 1/.style={sibling distance=30mm, level distance=1.0cm},
    level 2/.style={sibling distance=40mm, level distance=1.0cm},
    level 3/.style={sibling distance=20mm, level distance=1.0cm},
  every node/.style = {shape=rectangle, rounded corners,
    draw, align=center,
    top color=white, bottom color=blue!20}]
  \node {Gesture Detection}
    child { node {Gesture Classification}
      child { node {Pushes}
        child { node {FWD} }
        child { node {FWD-BWD} }}
      child { node {Swipes or Hold Hand} 
        child {node {Swipes} 
          child {node {LTR} }
          child {node {RTL} }}
        child {node {Hold Hand} }}}
     child { node {No Gesture} };
         
\end{tikzpicture}
\caption{Detection Classification Hierarchy}
\label{hierarchy}
\end{figure}

Support Vector Machine (SVM) classifier was chosen to detect and classify a predefined set of five gestures. SVM algorithm is proven to achieve a good performance for real world applications and with mathematical models that are based on simple ideas and are easy to analyze. 
Motion Frames obtained from consecutive blocks over time form a Motion Profile. On average a gesture occupies 2 seconds in time, generating a Motion Profile of 100 frames. Each gesture type has its unique Motion Profile. Two main features sets were engineered from the Motion Profile (100 frames length) to be used for detection and classification. The first set is RSS Vector, formed by summing the values of the highest 20 peaks over the profile length in each frame. The second set is Range Matrix, consisting of positions and values of the highest 20 peaks in each frame. Figure \ref{rss_range} shows the generation of the two feature sets.
Detection and classification is done in a hierarchy way as show in figure \ref{hierarchy}. The first two levels in the hierarchy are classified using RSS vector, and the last two levels are classified using Range Matrix.

\section{Experimental Evaluation}
\label{sec:majhead}
This section presents the experimental setup used for testing the system. A customized board with a mic and MEMS ultrasonic transmitter is connected to a PC through a sound-card. During experiments, data is recorded and saved to be processed off-line using MATLAB.\\ 
A set of around 120 repetitions of each type of the five gestures was  recorded. Test was done using cross validation where $8\%$ of the data was taken as a test set and the rest was taken as a training set. The following confusion matrix shows the performance of the system.

\begin{table}[h!]
\begin{center}
\begin{adjustbox}{max width=0.45\textwidth}
\begin{tabular}{| l | l | l | l | l | l |}
  \hline

{Gesture}   & {Right-Left} & {Left-Right} & {Hold Hand} & {Fwd-Bwd} & {Fwd} \\\hline
{Right-Left}      & \bm{$0.9423$}  & $0.0385$  & $0$  & $0$  & $0$\\\hline
{Left-Right}      & $0.0088$  & \bm{$0.9381$}  & $0.0885 $  & $0$  & $0$\\\hline
{Hold Hand}      & $0.0309$  & $0.2165$  & \bm{$0.8247$}  & $0$  & $0.0103$\\\hline
{Fwd-Bwd}      & $0.0085$  & $0.0169$  & $0.1356$  & \bm{$0.8136$}  & $0.1017$\\\hline
{Fwd}      & $0$  & $0.0545$  & $0.0545 $  & $0.0273$  & \bm{$0.9182$}\\\hline
   \end{tabular}
   \end{adjustbox}
\caption{Confusion matrix for classification results}
  \label{conf_mat}
\end{center}
\end{table}

\section{Conclusion}
In this paper we showed that a single ultrasonic transmitter and a single receiver can be used to detect and classify five different types of hand gestures with high accuracy. The design of the transmitted signal uses LFM utilizing a small bandwidth. Least squares SVM classifier with a polynomial kernel function was used to detect and classify gestures. System evaluation shows an average accuracy of $88.7\%$. Since the proposed system is using a single ultrasonic transmitter and a single receiver, it can be suitable for devices like laptops and mobile phones.\\

\bibliographystyle{IEEEbib}
\bibliography{IEEEbib}

\end{document}